\documentclass[10pt,letterpaper]{article}

\usepackage{opex3}

\usepackage[draft]{hyperref}

\usepackage{bm}
\usepackage{amsmath}
\usepackage{amssymb}

\usepackage{graphicx,color}
\usepackage{epsfig}

    \newcommand{\ket}[1]{\ensuremath{|\,{#1}\,\rangle}}
    \newcommand{\Sket}[1]{\ensuremath{\left|\,{#1}\,\right\rangle}}

    \newcommand{\lsub}[1]{\ensuremath{_{_{\!\scriptstyle #1}}}}

    \newcommand{\itgf}[1]{\ensuremath{\int\!\!d{#1}\,}}

    \newcommand{\sinc}{\ensuremath{\mbox{\hspace{1.3pt}sinc}\,}}

\begin{document}

\title{Manipulating spatial qudit states with programmable optical devices}

\author{G. Lima,$^{1,*}$ A. Vargas,$^{2}$ L. Neves,$^{1}$ R. Guzm\'{a}n,$^{2}$ and C. Saavedra$^{1}$}

\address{
$^1$Center for Quantum Optics and Quantum Information,
Departamento de F\'{\i}sica, Universidad de Concepci\'on,  Casilla 160-C, Concepci\'on, Chile.\\
$^2$Departamento de Ciencias F\'{\i}sicas, Universidad de La Frontera, Temuco, Casilla 54-D, Chile. \\
$^*$Corresponding author: glima@udec.cl}

\begin{abstract}
The study of how to generate high-dimensional quantum states (qudits) is justified by the advantages that they can bring for the field of quantum information. However, to have some real practical potential for quantum communication, these states must be also of simple manipulation. Spatial qudits states, which are generated by engineering the transverse momentum of the parametric down-converted photons, have been until now considered of hard manipulation. Nevertheless, we show in this work a simple technique for modifying these states. This technique is based on the use of programmable diffractive optical devices, that can act as spatial light modulators, to define the Hilbert space of these photons instead of pre-fabricated multi-slits.
\end{abstract}


\section{Introduction}

Liquid crystal displays (LCD) are programmable optical devices which can act as diffractive, reflective and refractive optical elements. This is possible because they are pixelated structures that can be used to produce spatially dependent amplitude and/or phase modulation of an incident light beam. These spatial light modulators (SLM) are used mostly for displaying images and to build digital lenses and holograms. However, because of their experimental versatility as programmable devices, their use has been slowly considered in the area of quantum optics. They have been used for implementing more advanced optical tweezers \cite{Twi1,Twi2}, for generating distinct types of quantum states \cite{Beck,Mark}, and as practical measurement instruments for the detection of the orbital angular momentum entanglement of the twin photons generated in the process of parametric down-conversion \cite{Zeilinger1,Padgett,Zeilinger2,India}.

In \cite{Zeilinger2} it was demonstrated that a SLM can also be used for creating and modifying entangled high-dimensional quantum states (qudits), whose Hilbert space is defined in terms of the orbital angular momentum of down-converted beams. High-dimensional quantum systems may have important applications. In quantum information, for example, the use of entangled qudits for communication would allow the transmission of more information per pair of particles shared, and more security against certain types of eavesdropping attacks \cite{Peres,Durt}. Besides, the study of entanglement between high-dimensional quantum systems, showed that entangled qudits violate locality stronger than qubits \cite{Zeilinger3} and, therefore, their use may help to close the detection loophole of Bell type experiments \cite{Bell,Aspect}. There have been several experimental investigations for the generation of photonic qudits. The biphoton polarization states of frequency-degenerate and frequency-nondegenerate parametric down-conversion were used to encode single-pure and -mixed states of qutrits ($D=3$) and ququarts ($D=4$) \cite{Kulik1,Mataloni,White2,Kulik2}. Other degrees of freedom which are suitable to generate qudits states are the time arrival of the twin photons \cite{Thew}, their longitudinal momentum \cite{Rossi}, and as we have demonstrated, recently, their linear transverse momentum, which can be explored for encoding states of at least eight-dimensional quantum systems \cite{Us}.

These qudits states are considered of hard manipulation and, therefore, have not being subject to exhaustive theoretical and experimental investigations to determine their real practical potential. In this work we follow the line of research initiated in \cite{Zeilinger2} to show that a SLM can be seen as ``... an active transformation in quantum optics experiments for the manipulation of photons coming from
spontaneous parametric down-conversion.'' We show that the coherent superposition of one spatial qudit state, which is generated by sending one of the down-converted beams through diffractive apertures, can be modified by considering the capability of amplitude-only modulation of the spatial light modulators, when it is used to define these type of apertures. This is because a spatial light modulator couples the transmitted photon spatial degree of freedom with its polarization, and this allows, through a projective measurement into the polarization Hilbert space, the control of the apertures transmissions which determine the real amplitudes of the spatial qudits states. Our work shows, therefore, that spatial light modulators allow the generation of new types of spatial qudits states, and also that they can be used for doing new projective measurements in the spatial qudits Hilbert space.

This paper is organized as follows: First, we give the mathematical description of the states that we generated in our experiment and then we describe how a SLM can be used to manipulate them. We also discuss, briefly, the calibration procedure to prepare the SLM for doing amplitude-only modulation of a down-converted beam. For doing this calibration we have considered the macroscopic technique proposed by Moreno \emph{et al.,} \cite{Moreno}, which does not require a previous calibration of the LCD internal microscopic parameters, and it is therefore interesting to see that it worked even when considering only coincidences counts in the calibration of the SLM, i.e, at the level of single photons. Finally, we show the results demonstrating the experimental generation and modification of the spatial qudits states done with the SLM.

\section{Experiment}

\subsection{Experimental Setup}

In spontaneous parametric down-conversion (SPDC) one photon from a pump beam incident onto a non-linear crystal originates, with small probability, two other photons usually called signal (s) and idler (i), twin photons, or also biphoton. In this process, the shape of the angular spectrum of the incident pump beam is transferred to the transverse amplitude of coincidence detection of the biphoton \cite{Monken}, and this allows one to control the transverse correlation properties of the down-converted fields by manipulating the pump field. By exploiting this effect, we have shown that it is possible to generate distinct types of two-photon states when diffractive optical elements are used for the transmission of the twin photons. Because the number of different ways available for their transmission defines their Hilbert space dimension, we call them spatial qudits. In \cite{Us}, we showed that maximally entangled states of spatial qudits can be generated by focusing the pump beam at the plane of pre-fabricated multi-slits placed on the propagation path of the twin-photons, and in \cite{ConcSpa,MixSpa}, we showed that other types of pure and mixed states of one and two spatial qudits can also be generated with these multi-slits, while considering distinct techniques for the manipulation of the pump field. Nevertheless, these manipulation strategies are complex and the family of the possible states which can be generated with them, is still very limited. This has motivated the development of other strategies to control the transverse correlation of the twin-photons \cite{Chinas,Peeters} and in this work we show that the use of spatial light modulators instead of pre-fabricated multi-slits to define the Hilbert space of the spatial qudit states, allows a technique for modifying these states which is simpler than the aforementioned techniques based on the manipulation of the pump beam.

\begin{figure}[t]
\centerline{\rotatebox{-90}{\includegraphics[width=0.45\textwidth]{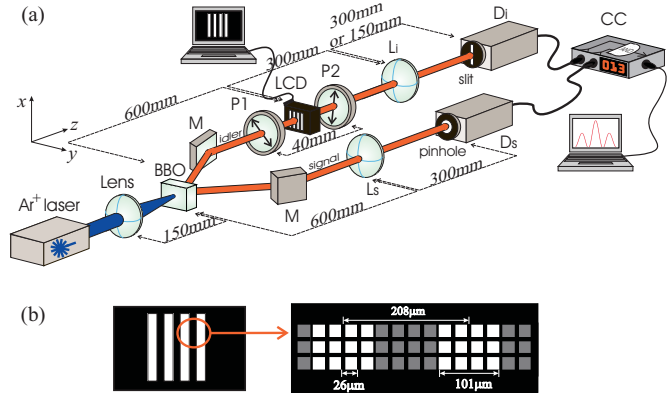}}}
\vspace{-0.6cm}
\caption{Schematic diagram of the experimental setup. See the main text for details. In this figure, BBO represents the non-linear crystal used to generate the twin photons. $L_i$ and $L_s$ are lenses with focal lengths of $150$~mm and $300$~mm, respectively, which are placed at the propagation path of the idler and the signal beams, at a distance of $900$~mm and $600$~mm of the BBO crystal, respectively. $P_1$ and $P_2$ are the polarizers which are placed close to a twisted nematic liquid crystal display (LCD) to build our spatial light modulator. The LCD panel is at a distance of $600$~mm of the BBO crystal and these polarizers are $2$~cm away from the LCD panel. The photo detectors of the idler and signal photons are represented by $D_i$ and $D_s$, respectively. CC is the circuit used to record the coincidence counts between these detectors. The idler photon is detected with the experimental setup in two distinct configurations: At the first one, the detector $D_{i}$ is placed at the LCD plane of image formation which is at $1200$~mm from the crystal. At the second one, this detector is moved to the focal plane of $L_{i}$ lens which is at $1050$~mm from the crystal. The signal photons are detected at the focal plane of lens $L_{s}$ which is at $900$~mm from the crystal. In (b) a sketch of the four-slit addressed on the LCD and its pixelated structure is shown.} \label{Fig:Setup}
\end{figure}

The experimental setup considered is outlined in Fig.~\ref{Fig:Setup}. A single-mode Ar$^{+}$-ion laser operating at 351.1~nm and with an average power of 150~mw is focused into a 5-mm-long $\beta$-barium borate crystal cut for type-II parametric down-conversion luminescence. Degenerated down-converted photons with the wavelength of 702~nm are selected using interference filters with small bandwidths ($10$~nm FWHM). The LCD is placed at the propagation path of the idler photon at a distance of $600$~mm from the crystal. Our SLM is composed of two polarizers ($P1$ and $P2$) and a twisted nematic liquid crystal display (CRL-Opto, model XGA2) with a spatial resolution of $1024\times768$ monochrome pixels. The polarizers are placed close to the LCD panel ($2$~cm away from each side). The LCD panel is connected to an interface PCB and it allows us to drive the programmable optoelectronics device from a computer. The horizontal and vertical pixel dimensions are 23~$\mu$m and 16~$\mu$m, respectively. Furthermore, the separations between pixels are 3~$\mu$m in the horizontal and 10~$\mu$m in the vertical directions. Initially, the SLM is addressed with a digital four-slit. The slit's width is $2a=101$~$\mu$m and the distance between two consecutive slits is $d=208$~$\mu$m (See Fig.~\ref{Fig:Setup}(b)). After being transmitted by the SLM, the idler photon propagates through a $150$~mm focal length lens ($L_{i}$), which is $300$~mm from the LCD. The idler photon is detected with the experimental setup in two distinct configurations. The first one is used for measuring the image of the aperture mounted in the LCD, and the second one for measuring the interference pattern of the transmitted idler beam at the focal plane of lens $L_{i}$. At the first configuration, the avalanche photodiode detector ($D_{i}$) is placed at the LCD plane of image formation which is at $1200$~mm from the crystal. At the second configuration, this detector is moved to the focal plane of $L_{i}$ lens which is at $1050$~mm from the crystal. The spatial mode of the signal photon is defined by pinholes placed along its propagation path. After the transmission through these pinholes, the signal photons are then focused with the lens $L_{s}$ inside the detector $D_{s}$. $L_{s}$ has a focal length of $300$~mm and it is placed at $600$~mm from the crystal. $D_s$ is kept fixed at the distance of $900$~mm from the crystal. The photo detectors $D_{i}$ and $D_{s}$ are connected to a circuit used to record the singles and coincidences counts.

Under such experimental configuration, the twin photon state in the plane defined by the SLM can be written as \cite{Us,Monken,Geraldo}
\begin{equation} \label{State}
\ket{\Psi} = \left(\sum_{l=-\frac{3}{2}}^{\frac{3}{2}}
          \alpha_l \ket{l}\lsub{i}\right) \otimes \ket{\gamma}\lsub{s},
\end{equation} where $\ket{\gamma}\lsub{s}$ is the spatial mode of the signal photon defined
by the pinholes of its propagation path and the state $\ket{l}\lsub{i}$ is a single photon state defined, up to a global phase factor, as
\begin{equation}      \label{base}
\ket{l}\lsub{i} \equiv \sqrt{\frac{a}{\pi}} \itgf{q_{i}}
e^{- iq_{i}ld}\sinc(q_{i}a)\ket{1q_{i}}.
\end{equation} It represents the state of the photon in mode $i$ transmitted by the slit $l$ of the SLM. $\ket{1q_{i}}$ is the Fock state of one photon in mode $i$ (photon idler) with the transverse wave vector $\vec{q}_{i}$. The index $l=\frac{-3}{2},\frac{-1}{2},\frac{1}{2},\frac{3}{2}$ stands for the first, second, third and fourth slit of the modulator, respectively. These states form an orthonormal basis in the four-dimensional Hilbert space of the idler photon and are therefore used to define the logical spatial qudits states. The coefficients $\alpha_l$ are in this case real functions which are related with the slits transmissions by \protect{$\alpha_l=\sqrt{\frac{t_l}{\sum t_l}}$}, where $t_l$ is the transmission of the slit-$l$.

Therefore, in our experiment, the signal photon is a trigger for the detection of the idler photon used to encode the four-dimensional spatial qudit state. The non-linear crystal works like a heralded single photon source and by monitoring the coincidence counts, we can see how the idler-qudit state is being modified by the SLM. From Eq.~(\ref{State}), one can see that the capability of the SLM for doing amplitude-modulation can be explored to change, independently, each slit transmission of the multi-slit addressed on it and thus, to modify the initial spatial qudit state superposition.

\subsection{Amplitude-only modulation of a down-converted beam}

The liquid crystal spatial light modulator must be calibrated for doing amplitude only or phase only modulation. In the literature, several works have reported optical configurations and methods for doing this calibration \cite{Moreno,Davis,Mogensen,Coy,Nicolas,Marquez}. One simple technique is the method based on the Jones matrix of the polarized states generated by the LCD. This is a macroscopic model which considers the SLM as a black box. By using only one single wavelength incident light beam it is possible to determine the Jones Matrix of the LCD. The predictive capability obtained after the reconstruction of the SLM Jones matrix allows one to obtain the optical configuration for doing the amplitude-only modulation with it. Since just one single wavelength light beam is necessary for implementing this protocol, we could use the down-converted beams and the coincidence counts to calibrate the modulator precisely for the wavelength of 702~nm. This has been done by illuminating the SLM with the idler photon and detecting the signal and the idler in coincidence as we described above.

\begin{figure}[tbh]
\vspace{-3cm}
\centerline{\includegraphics[width=0.7\textwidth]{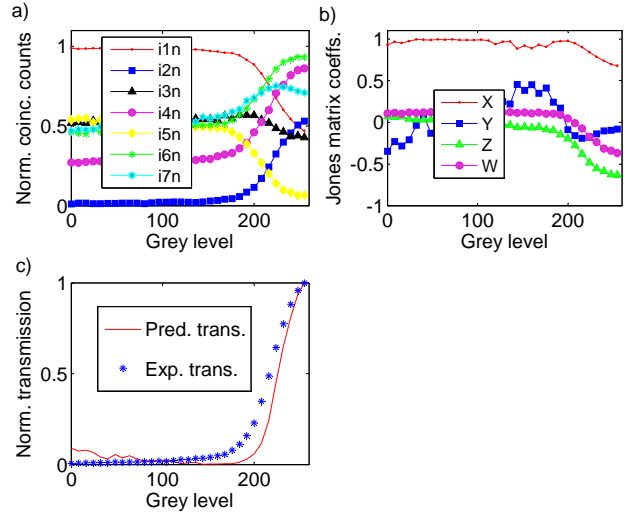}}
\vspace{-2.8cm}
\caption{In (a) it is shown the seven irradiance measurements performed to determine our SLM Jones matrix coefficients. i1n corresponds to an optical configuration where the first and the second polarizers were aligned at the horizontal direction ($P1 = P2 = H$). For the others curves their directions were: i2n: $P1=H$ and $P2=V$; i3n: $P1=H$ and $P2=L$; i4n: $P1=R$ and $P2=V$; i5n: $P1=45^{\circ}$ and $P2=H$; i5n: $P1=45^{\circ}$ and $P2=V$; i7n: $P1=H$ and $P2=45^{\circ}$, where the indexes H, V, R, L and $45^{\circ}$ stand for the horizontal, vertical, right circular, left circular and diagonal polarization directions, respectively. In these curves, the integration time of the data points recorded was $5$~s. In (b) we have the values of the Jones matrix coefficients $X$, $Y$, $Z$ and $W$ of our SLM. In (c) the predicted and the experimental curves for the transmission of the SLM as a function of its grey level are shown.} \label{Fig:Calibr}
\end{figure}

The Jones matrix of non-absorbing polarization devices is unitary and depends on four real-valued parameters, usually called $X$, $Y$, $Z$ and $W$ \cite{Moreno,Pousa}
\begin{equation}
M = \exp(-i \varpi) \left[\begin{array}{cc}
X -i Y &Z-iW \\
-Z-iW &X+iY \\
\end{array}\right],
\end{equation} where $\varpi$ is a global phase shift. The method of Moreno \emph{et. al} \cite{Moreno}, determines the values of these parameters for a liquid crystal display through seven irradiance measurements, which are recorded after the transmission of the light through the SLM and by varying its grey level and also its optical configuration. In Fig.~\ref{Fig:Calibr}(a), we can see the normalized coincidence counts recorded as a function of the grey level of our SLM for these measurements. In this figure the curve $i1n$ corresponds, for example, to the optical configuration at which the initial and the second polarizers of the SLM are set to the linear horizontal polarization direction. In Fig.~\ref{Fig:Calibr}(b), we can see the values for the Jones matrix coefficients of our SLM as a function of its grey level. In Fig.~\ref{Fig:Calibr}(c), the predicted and the experimental curves for the transmission of the SLM, as a function of its grey level, are shown for a specific optical configuration, where our SLM is supposed to be modulating only the amplitude of a incident light at the wavelength of 702~nm. In this configuration, the phase being modulated should be constant regardless of the grey level that the pixels of the liquid crystal display are set for.

\subsection{Results}

\begin{figure}[tbh]
\vspace{3cm}
\begin{center}
\includegraphics[width=0.5\textwidth]{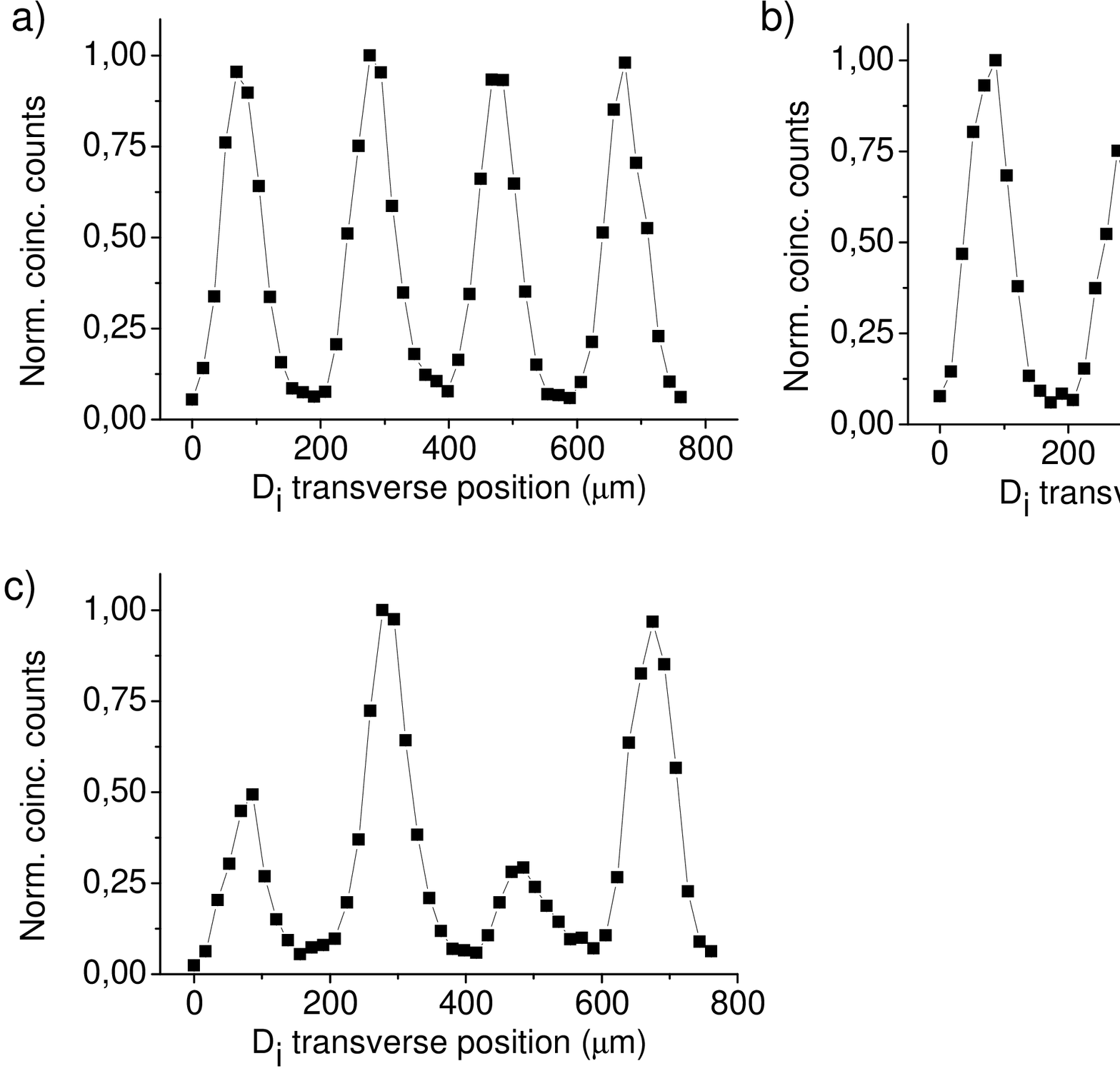}
\end{center}
\vspace{-2.5cm}
\caption{Normalized coincidence counts recorded with detector $D_i$ scanning transversally to the image of the SLM-four-slits. The coincidences counts were recorded with a integration time of $5$~s. In (a) it is shown the image of the initial four-slit [100, 100, 100, 100]. In (b) we have the image recorded for the modified four-slit [100, 75, 50, 25] and in (c) the image of the four-slit [50, 100, 25, 100]. See the text for details.} \label{Fig:Im}
\end{figure}

\begin{figure}[tbh]
\vspace{3cm}
\begin{center}
\includegraphics[width=0.6\textwidth]{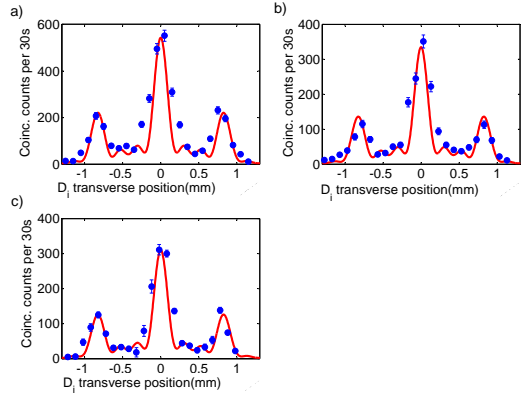}
\end{center}
\vspace{-3cm}
\caption{Interference patterns recorded in coincidence as a function of the detector $D_i$ transverse position. In (a) it is shown the interference of the initial four-slit [100, 100, 100, 100]. In (b) it is shown the interference pattern of the modified four-slit [100, 75, 50, 25]. In (c) it is shown the interference pattern of the modified four-slit [50, 100, 25, 100]. See the text for details. The solid lines were obtained theoretically considering the states of Eq.~(\ref{qudit1}), Eq.~(\ref{qudit2}) and Eq.~(\ref{qudit3}), respectively.} \label{Fig:Int}
\end{figure}

As we discussed above, initially, our SLM is addressed with a four-slit whose slits transmission are equal to $100\%$ ([100, 100, 100, 100]). According to Eq.~(\ref{State}), the qudit state being encoded in the transmitted idler photon is given by
\begin{equation} \label{qudit1}
\ket{\zeta_1}\lsub{i}=\frac{1}{2}\times \left( \Sket{\frac{-3}{2}} + \Sket{\frac{-1}{2}} + \Sket{\frac{1}{2}} + \Sket{\frac{3}{2}} \right).
\end{equation}  As we have demonstrated in \cite{GLima}, the real part of the amplitudes of the qudit states can be checked with coincidence measurements performed at the image plane of the diffractive apertures used to define these qudits. The measurement that we have performed at the plane of this SLM image is shown in Fig.~\ref{Fig:Im}(a). According to Eq.~(\ref{qudit1}), the coincidence peaks which appear as the detector $D_i$ is scanned transversely to the SLM-multi-slit image should be equally weighted. One can clearly see the good agreement between our theory and the obtained experimental results. The measurements performed at the image plane, however, do not guarantee that the state of the SLM-qudit can indeed be represented by a coherent superposition of the logical spatial states as given by Eq.~(\ref{qudit1}). There are other distinct types of states that would generate the same results; for example: a pure state where the amplitudes of the superposition are complex and also a completely incoherent mixed state. Therefore, it is necessary to implement another type of measurement which exclude these possibilities. This is done by measuring the coincidence interference pattern formed at the focal plane of the $L_i$ lens. For doing this measurement, the idler detector is now moved to a distance of $1050$~mm from the crystal which corresponds to the focal plane of $L_i$. For a pure state whose amplitudes have distinct phases, the maximal peak of this pattern would be shifted out of the transverse center of the SLM-multi-slit. If the state is completely incoherent, there will be no interference at all. The coincidence interference pattern recorded for the four-slit [100, 100, 100, 100] is shown in Fig.~\ref{Fig:Int}(a). One can clearly see the interference fringe structure without any phase shift. The solid curve represents the detection probability calculated by considering the state of Eq.~(\ref{qudit1}) and a extra term which accounts for the finite size of the source. This reduces the idler-degree of transverse coherence at the LCD-plane and explain the less-than-one visibility of the patterns observed. The good fit between our experimental results and this theoretical curve means that Eq.~(\ref{qudit1}) can indeed be seen as a good approximation for the experimentally generated initial state.

To show that spatial light modulators can be used to manipulate this spatial qudit state, we used the amplitude modulation of the SLM to change the slits transmissions of the previously addressed four-slit, used to define this initial qudit state. To modify, independently, the transmission of each slit, one just need to change the grey level of the pixels used to form the slits apertures (See Fig.~\ref{Fig:Setup}(b)). We have considered two new four-slits with their slits set with different grey levels. In the first one the slits transmissions are [100, 75, 50, 25] percent. On the second one they are [50, 100, 25, 100] percent. The modified spatial qudit states should now be given by
\begin{equation} \label{qudit2}
\ket{\zeta_2}\lsub{i}= 0.63 \Sket{\frac{-3}{2}} + 0.54 \Sket{\frac{-1}{2}} + 0.44 \Sket{\frac{1}{2}} + 0.31 \Sket{\frac{3}{2}},
\end{equation} and
\begin{equation} \label{qudit3}
\ket{\zeta_3}\lsub{i}= 0.42 \Sket{\frac{-3}{2}} + 0.60 \Sket{\frac{-1}{2}} + 0.30 \Sket{\frac{1}{2}} + 0.60 \Sket{\frac{3}{2}},
\end{equation} respectively.

To confirm these superposition as good approximations for the experimentally modified states, we measured again at the image plane of these new four-slits. The coincidence peaks should be equivalents to the slits transmissions of these apertures. The results for $\ket{\zeta_2}\lsub{i}$ and for $\ket{\zeta_3}\lsub{i}$ are shown in Fig.~\ref{Fig:Im}(b) and Fig.~\ref{Fig:Im}(c), respectively. In order to verify that the amplitude-modulation of the slits with different grey levels was not introducing complex phases or decoherence to the state, we measured again the interference patterns of the these new four-slits. The patterns recorded with [100, 75, 50, 25] and with [50, 100, 25, 100] are shown in Fig.~\ref{Fig:Int}(b) and Fig.~\ref{Fig:Int}(c), respectively. Again we have the coincidence interference patterns without phase shifts. These curves also fit well with the theoretical predictions for the detection probabilities calculated in the same way explained before, but now considering Eq.~(\ref{qudit2}) and Eq.~(\ref{qudit3}) for the qudit superpositions. These results, together, demonstrate that the amplitude-only modulation of programmable spatial light modulators can indeed be used for modifying, in a controlled way, the spatial qudit states.

\section{Conclusion}

In this work we demonstrated that spatial light modulators based on liquid crystal displays can be used to manipulate the spatial qudit states. The technique here discussed is simpler than previous ones used to modify these states. It can also be used after their generation. Therefore, it can be used for example, to concentrate the entanglement between propagated entangled spatial qudits states. The ability for doing simple manipulations of a quantum state is a prerequisite for implementing quantum protocols with it. The work here is, therefore, relevant for the field of quantum information where qudits states can bring significative advantages. Besides this, we also envisage the use of spatial light modulators to implement optimal techniques of state estimation for the spatial qudit states \cite{Tomospa,Japas}. In the case where the dimension of the qudit is a prime or a power of prime, for example, the detection of the down-converted photon in mutually unbiased bases is the optimal choice for implementing its quantum state tomography \cite{Klimov}. These mutually unbiased bases can be generated for the spatial qudits by using the SLM to generate conditional phase-shifts between the slit-states. This can be done by exploiting its capability of also doing phase-modulation. Another possibility is the use of the SLM amplitude and phase modulation for implementing logical gates for the spatial qudits.

\section*{Acknowledgments}

This work was supported by Grants Milenio ICM P06-067F and FONDECYT 110850555. A. Vargas acknowledges to Convenio de Desempe\~{n}o de la Universidad de La Frontera. R. Guzm\'{a}n thanks to CONICYT-PBCT Red21.

\end{document}